# Evaluating Medical IoT (MIoT) Device Security using NISTIR-8228 Expectations

Thomas P. Dover[1]

**ABSTRACT**

How do healthcare organizations (from small Practices to large HDOs[2]) evaluate adherence to the cybersecurity and privacy protection of Medical Internet of Things (MIoT)[3] used in clinical settings? This paper suggests an approach for such evaluation using National Institute of Standards and Technology (NIST) guidance. Through application of NISTIR[4] 8228 *Expectations* it is possible to quantitatively assess cybersecurity and privacy protection, and determine relative compliance with recommended standards. This approach allows organizations to evaluate the level of risk a MiOT device poses to IT systems and to determine whether or not to permit its use in healthcare/IT environments.

This paper reviews the current state of IoT/MiOT cybersecurity and privacy protection using historical and current industry guidance & best-practices; recommendations by federal agencies; NIST publications; Executive Orders (EO) and federal law. It then presents similarities and differences between IOT/MiOT devices and "traditional" (or classic) Information Technology (IT) hardware, and cites several challenges IoT/MiOT pose to cybersecurity and privacy protection.

Finally, a practical approach to evaluating cybersecurity and privacy protection is offered along with enhancements for validating assessment results. In so doing it will demonstrate general compliance with both NIST guidance and HIPAA/HITECH requirements.

---

[1] Adjunct Faculty, Butler County Community College, Butler, PA. Email: thomas.dover@bc3.edu

[2] Health Delivery Organization (HDO)

[3] aka Internet of Medical Things (IoMT) or Internet of Healthcare Things (IoHT).

[4] NISTIR – 'NIST Internal Report'. IRs differ from Special Publications (SP) in that IRs represent work performed by NIST for outside sponsors (government or non-government).
Source: https://csrc.nist.gov/glossary/term/National_Institute_of_Standards_and_Technology_Interagency_or_Internal_Report

## TARGET AUDIENCE

This paper is directed towards organizations or businesses whose information systems are responsible for providing service to, or otherwise supporting, healthcare providers, healthcare delivery organizations or any other entity which creates, handles, stores or transmits Electronic Protected Health Information (ePHI), and whose information or network system(s) employ MIoT devices for this purpose[5].

## SCOPE

This paper is limited to cybersecurity and privacy protection assessment & evaluation of MiOT devices used by the healthcare sector as they relate to HIPAA[6] and HITECH[7] requirements for protection of electronic Protected Health Information (ePHI).

**U.S. Critical Infrastructure Sector:** Healthcare and Public Health Sector
**Responsible Department:** Health and Human Services (HHS)
**HHS Agencies:** Office of Civil Rights (OCR), Centers for Medicare & Medicaid Services (CMS), Food and Drug Administration (FDA)

## INTRODUCTION

According to Cisco Internet Business Solutions Group the 'Internet of Things' (IoT) began sometime between 2008 and 2009 when the number of "things or objects" connected to the internet exceeded the number of people connected.

Published references to *Medical Internet of Things* (MIoT) most likely started between 2012 and 2013. In 2012, the Government Accounting Office (GAO) recommended in the August edition of its *Highlights* report to Congress[8] that the FDA should "develop and implement a plan expanding its focus on information security risks." Indeed, in 2013[9] the Food and Drug Administration (FDA) issued medical device manufacturers guidance for the cybersecurity of medical IoT devices[10] which represented the agency's "current thinking on this topic. [3]"

The Healthcare sector has been quick to incorporate MIoT into clinical operations as such use offers greater efficiency, improved operations, cost savings and most importantly, improved patient outcomes. Examples where MIoT are employed include blood pressure and glucose level monitoring, pulse oxymeters, weight/BMI scales, thermometers, spirometers, and EKG monitoring.

The key to success of MIoT (in fact, all IoT) is internet-connectivity and the ability of MIoT devices to transmit (patient) information.

---

[5] It should be emphasized that although the assessment approach presented in this paper is specific to Healthcare it can be applied to any industry or sector where IoT is employed.

[6] Health Insurance Portability and Accountability Act (HIPAA). 1996.
[7] Health Information Technology for Economic and Clinical Health (HITECH). 2009.
[8] GAO Highlights, GAO-12-816.
[9] *Content of Premarket Submissions for Management of Cybersecurity in Medical Devices*. Guidance for Industry and Food and Drug Administration Staff. First draft published June 14, 2013. Food and Drug Administration, et al.
[10] It should be noted that the FDA issued guidance for "…Software Contained in Medical Devices" in May, 2005, however, this publication pre-dated iOT and concerned itself with embedded software.

In December 2020, President Trump signed into law the *IoT Cybersecurity Improvement Act of 2020*. This law, in part, directs the National Institute of Standards and Technology (NIST) to take steps for the increased cybersecurity of IoT. In accordance with this law, in late December NIST published Special Publication (SP) 800-213 (*IoT Device Cybersecurity Guidance for the Federal Government: Establishing IoT Device Cybersecurity Requirements*). Although written for government agencies guidance provided by NIST Special Publications can be used by non-government organizations as well and SP.800-213 is no exception.

On May 12, 2021, President Biden issued an *Executive Order on Improving the Nation's Cybersecurity.* This EO includes a provision for 'Enhancing Software Supply Chain Security'. This requirement, in part, calls for the development of a "Software Bill of Materials"; a list of components used by developers or manufacturers to build software applications. Such a document would be invaluable for evaluating MIoT security.

Based on NIST guidance, a qualitative framework for assessing IoT (and by extension MIoT) cybersecurity and privacy protection is possible. Using *Expectations* for MIoT cybersecurity and privacy protection outlined in NIST.IR 8228 (and associated publications[11]) a set of security and privacy criteria is used to assess compliance, and in turn, evaluate risk that a MIoT device may pose to a healthcare organization's IT environment. Moreover, such evaluation is crucial for complying with HIPAA/HITECH regulations governing the Confidentiality, Integrity and Availability (CIA) of Protected Health Information (PHI).

## MIoT GOVERNANCE & OVERSIGHT

By definition, MIoT devices are IoT devices used for a specialized purpose (healthcare). Specialization notwithstanding, MIoT devices are subject to the same cybersecurity and privacy protection requirements as non-MIoT devices. According to the Food and Drug Administration (FDA), cybersecurity is a shared responsibility between the FDA and "device manufacturers, hospitals, healthcare providers, patients, security researchers, and other government agencies including the U.S. Department of Homeland Security's Cybersecurity & Infrastructure Security Agency (CISA) and U.S. Department of Commerce [20]."

## U. S. FOOD AND DRUG ADMINISTRATION (FDA)

The Department of Health and Human Services (HHS), Food and Drug Administration (FDA) is responsible for medical device oversight. According to the General Accounting Office (GAO), the FDA is responsible "for ensuring the safety and effectiveness of medical devices in the United States" [3].

In 2012, the GAO recommended in its August *Highlights* report to Congress[12] that the FDA should "develop and implement a plan expanding its focus on information security risks."

Pursuant to its responsibility the FDA has published guidance for both Premarket (2014) and Postmarket (2016) management of cybersecurity in MiOT devices.

[11] NIST Cybersecurity Framework (CSF), NISTIR 8259 series, SP.800-53r5 (Security Control Catalog) and SP.800-213 (IoT Device Security Guidance for the Federal Government).
[12] GAO Highlights, GAO-12-816.

In 2014, FDA Center for Devices and Radiological Health, released *Content of Pre-Market Submissions for the Management of Cybersecurity in Medical Devices* (FDA 1825). This publication provides guidance for Industry, and FDA staff. Though 'Internet of Things' is not specifically mentioned in the document--understandable given that MiOT devices were in the very early stages of being applied to the Healthcare sector—it nevertheless recommends that "medical device manufacturers address cybersecurity during the design and development of the medical device, as this can result in more robust and efficient mitigation of patient risks" [1]. Cybersecurity areas addressed included identification of threats and vulnerabilities; assessment of the impact of threats on device functionality and end users\patients; assessment of the likelihood of threat\vulnerability occurring; determination of risk levels; and assessment of residual risk and risk acceptance criteria. Moreover, the recommendations specifically cite NIST Cybersecurity Framework categories *Identify*, *Protect*, *Detect*, *Respond* and *Recover*.

In 2016, FDA Center for Devices and Radiological Health, released *Postmarket Management of Cybersecurity in Medical Devices*. Similar to FDA 1825, this publication clarifies postmarket recommendations for manufacturers to follow relative to identifying, monitoring and addressing cybersecurity vulnerabilities and exploits as "part of their postmarket management of medical devices" [18].

**NATIONAL INSTITUTE OF STANDARDS AND TECHNOLOGY (NIST) GUIDANCE**

In 2019, the National Institute of Standards and Technology released Internal Report[13] (IR) NISTIR 8228, *Considerations for Managing Internet of Things (IoT) Cybersecurity and Privacy Risks*.

In 2020, NISTIRs 8259(A)(B)(C)(D) were released as supplementary/complementary Guides for IoT cybersecurity. These publications provided guidance to IoT device manufacturers (8259/8259A); guidance for non-technical support capabilities (8259B); guidance for Core security baselines (8259C); and guidance for creating a Profile for IoT Baselines (for the federal government) (8259D).

Pursuant to passage of federal law governing IoT cybersecurity[14], in December, 2020 NIST published SP.800-213, *IoT Device Cybersecurity Guidance for the Federal Government*. This publication outlines IoT device cybersecurity requirements and references several NIST Guides which have cross-application to HIPAA and HITECH requirements. Among them: NIST Cybersecurity Framework, SP.800-53r5 (Security Control Catalog), NISTR-8228 and the NISTIR 8259 series. Collectively, these Guides provide the basis for a framework which can be used to determine MIoT compliance with cybersecurity and privacy protection.

While SP.800-213 asks useful questions--labeled as "Cybersecurity Considerations"--which certainly aid in evaluating the security of an IoT device they are broad in scope and therefore not suitable for a tailored security assessment. For example, "What is the benefit of the IoT device and how will it be utilized? [11]" can provide direction for more targeted questions. The ten (10) questions listed in SP.800-213 may therefore be better suited for general summary.

[13] NISTIR (Internal or Interagency Report). Reports of research findings, including background information for FIPs and SPs. Source: https://csrc.nist.gov/publications/. Retrieved: 03/22/21.
[14] Internet of Things Cybersecurity Act of 2020. Signed into law on December 04, 2020.

In addition to FDA and NIST guidance the private sector has added its perspective to MiOT cybersecurity and privacy protection. In 2018, the Medical Device Innovation Consortium[15] (MDIC), a non-profit, public-private partnership published *Medical Device Cybersecurity Report*: *Advancing Coordinated Vulnerability Disclosure* with the purpose of "coordinated vulnerability disclosure (CVD) policies by medical device manufacturers" [8].

**EXECUTIVE ORDER (EO)**

On May 12, 2021, President Biden issued (an) *Executive Order on Improving the Nation's Cybersecurity* [9]. This intention of this EO is to "improve the nation's cybersecurity and protect federal government networks." The EO addressed several areas of cybersecurity including barriers to sharing information; modernizing cybersecurity standards; improving software supply chain security; creating a standard playbook for cyber-incidents and other areas.

As stated in the EO, Software Supply Chain Security guidelines

> include criteria that can be used to evaluate software security, include criteria to evaluate the security practices of the developers and suppliers themselves, and identify innovative tools or methods to demonstrate conformance with secure practices.

Guidelines also call for the development of a Software Bill of Materials (SBOM). As defined in the EO, an SBOM is a "formal record containing the details and supply chain relationships of various components used in building software."

Though Executive Orders only apply to federal agencies their effect tends to influence the behavior of non-government organizations (NGO) who either do business with the federal government or are impacted by its rules and regulations. NGOs will endeavor to either satisfy specific requirements or otherwise adhere to EO mandates. It is likely, therefore, that developers of software either embedded within or used by MiOT devices will work to create SBOMs for their products/services.

**FEDERAL LAWS AFFECTING MiOT**

New laws were enacted in 2020 and 2021 that will have an important impact on MiOT cybersecurity and privacy protection.

On December 4, 2020, the *IoT Cybersecurity Improvement Act of 2020*[16] was signed into law by President Trump. This law required, in part, that NIST develop standards and guidelines for the federal government to follow governing IoT devices used or controlled by a government agency. As stated earlier, NIST guidance can be readily adapted by private-sector companies and organizations.

---

[15] https://mdic.org/

[16] H.R. 1668, Public Law 116-207. The IoT Cybersecurity Act of 2020 was first introduced into Congress in 2017.

On January 5, 2021, President Trump signed into law an amendment to the Health Information Technology for Economic and Clinical Health Act (HITECH). HR 7898, otherwise known as the *HIPAA Safe Harbor* provision directs HHS to factor, in part, an organization's use of industry-standard cybersecurity practices during the previous twelve (12) months when investigating suspected data breaches or other violations. The amendment is meant to encourage healthcare providers and organizations to use NIST best-practices when formulating their cybersecurity and privacy protection strategies.

## CHALLENGES OF MIoT IN HEALTHCARE

The Healthcare sector is governed by the provisions of HIPAA and HITECH. Both contain specific regulations and\or requirements for the protection of ePHI and other sensitive information. Any process, system or device used to create, transmit or store ePHI is subject to these provisions.

For healthcare organizations and providers MIoT devices represent several cybersecurity and privacy protection challenges. Central to which is that MIoT devices do not behave, operate, or perform in the same manner as traditional[17] IT devices. This difference is due to a MIoT device's core functions *Sensing* (retrieving and transmitting information about the real world and transmitting it) and *Actuating* (making changes to the physical world). Some of the differences between MIoT and traditional IT include:

1. The ability to configure, update and monitor
2. Lack of transparency (black box problem)
3. Compatibility with existing Infrastructure
4. Information security (CIA)
5. Third-party access

**The ability to configure, update and monitor** means, at a minimum, having access to the MIoT device in order to perform routine and as-needed management functions such as access control (e.g., passwords), software updates (i.e., patch management) and log review. Due to manufacture design and production this level of access may not be available or even possible. It may not even be possible to know, to a reasonable degree of certainty, if the MIoT device is functioning properly or at all.

**Lack of transparency (black box problem)** is an issue with some MIoT devices due to their design and manufacture. Such devices do not allow insight into their configuration, operational settings or performance/activity logs. Lack of transparency prevents normal or routine cybersecurity and privacy protection oversight, and introduces risk into an IT environment since the state of compliance (with HIPAA/HITECH regulations) is unknowable.

**Compatibility with existing Infrastructure** may cause concern for established Datacenters and IT networks. Since MIoT devices may operate and function differently than traditional IT devices such difference can result in incompatibility with systems that were not designed for MIoT integration. In turn, MIoT devices may require new management systems for proper operation and oversight with IT Departments finding it necessary to add resources (staff & skills or external services) to manage MIoT deployment within their networks.

[17] or "classic" IT devices such as routers, switches, servers, etc…

**Information Security (CIA)** is a core tenant of HIPAA. CIA means taking appropriate steps to protect the Confidentiality, Integrity and Availability of protected health information (PHI). If a MIoT device stores data (not all do) it is critical that it be protected—depending on whether or not said information is PHI or CUI[18]--using cryptography or some other means of data protection. 'Black-box' devices or devices without any type of access or insight into their operation or status introduces significant risk in the event of exploit or compromise.

**Third-party access** is of concern for MIoT devices which permit no end-user access to their configuration settings or operational status (only third-parties). Such "unmanaged" devices may prevent real-time access during operational error or failure with access delayed further if the third-party is unavailable. It may also prevent those responsible from properly gauging network or operational status of a MIoT device when a software or firmware update is required, device EOL[19] is reached, or for other routine management and maintenance functions. In addition, a manufacturers Software Bill of Materials (SBOM) may be unavailable to a healthcare provider that is considering using MiOT devices in its clinical setting(s).

## USING NISTR 8228 'EXPECTATIONS' FOR MIoT CYBERSECURITY ASSESSMENT

Using NIST guidance (augmented by federal law), FDA guidance and private-sector recommendations a simple framework can be created for assessing the level of cybersecurity and privacy protection a MIoT device possesses. By using the framework the level of overall compliance (with recommended standards) can be obtained which, in turn, can be used to determine acceptable risk.

NISTIR 8228 presents three high-level challenges to IoT cybersecurity and privacy protection; *Protect Device Security*, *Protect Data Security*, and *Protect Individuals Privacy.* Within each are sub-goals such as Asset Management and Vulnerability Management. *Expectations* are used to identify specific properties or capabilities an IoT device must possess in order to ensure proper cybersecurity and privacy protection (Figure 1). In addition, references to NIST Cybersecurity Framework and NIST.SP 800-53r5 (Security Control Catalog) are provided along with organizational Implications.

[18] Confidential but Unclassified Information (CUI). Reference NIST.SP-800-171 & NIST.SP-800-172.
[19] end-of-life (EOL)

**Table 1: Potential Challenges with Achieving Goal 1, Protect Device Security**

| Challenges for Individual IoT Devices, and Risk Considerations Causing the Challenges | Affected Draft NIST SP 800-53 Revision 5 Controls | Implications for the Organization | Affected Cybersecurity Framework Subcategories |
|---|---|---|---|
| **Asset Management** | | | |
| Expectation 1: The device has a built-in unique identifier. | | | |
| 1. The IoT device may not have a unique identifier that the organization's asset management system can access or understand.<br>Risk Consideration 2 | • CM-8, System Component Inventory | • May complicate device management, including remote access and vulnerability management. | • ID.AM-1: Physical devices and systems within the organization are inventoried |

Figure 1

By assessing individual compliance with twenty-five (25) *Expectations* overall compliance can be assessed and quantified. This information can then be used to a) evaluate risk, b) identify MiOT weaknesses or vulnerabilities, c) gauge the level of IT management required and d) evaluate whether or not to allow the MiOT device into a network.

In addition to NISTIR 8228 *Expectations*, associated NISTIR 8259 series publications identified three (3) cybersecurity requirements not contained in NISTIR 8228. Since NISTIR 8259 series is intended for IoT manufacturers or federal government agencies these additional requirements are included in the assessment framework as optional considerations since it may not be possible to obtain sufficient detail to satisfy their security requirement(s).

## NISTIR 8228 SECURITY EVALUATION PROCESS

The evaluation process consists of determining compliance (with NISTIR 8228 *Expectations*) and assessing the level of risk (acceptable, correctable, unacceptable) an MiOT device presents to a healthcare organization's IT environment. The process should also include evaluation with HIPAA/HITECH requirements for the protection of PHI. Compliance-with-*Expectation* should be completed by either the MiOT device manufacturer or vendor on behalf of the healthcare provider considering its use. An assessment can be completed by the healthcare provider but may be inaccurate—through no fault of the healthcare provider--unless supporting or corroborating evidence (or documentation) is provided by manufacturer or vendor.

## MIoT SECURITY ASSESSMENT WORKBOOK\TOOL

Microsoft Excel was used to create the MiOT Security Assessment Workbook (Figure 2). The workbook utilizes NISTIR 8228 *Expectations* to establish a quantitative framework for assessing MiOT cybersecurity and privacy protection compliance. In addition to specific *Expectation* requirements the workbook provides references to associated NIST publications for each requirement.

**Note:** the Excel workbook is not included as part of this paper but can be obtained by contacting the author[20].

[20] thomas.dover@bc3.edu

ORGANIZATION NAME

Medical Internet-of-Things (IoT) - Cybersecurity Survey

**Directions:** answer the question using the "Compliance" field ("Value" field is calculated based on answer).

Reference NIST.IR-8228 (published June, 2019), et al.
*Compliance key at bottom of spreadsheet
** Compliance with HIPAA Security Rule

| # | DEVICE SECURITY | Compliance* | Value | Validation Point/Tool | Control Type** | Comments | N I Cybersecurity Framew |
|---|---|---|---|---|---|---|---|
| | ASSET MANAGEMENT | | | | | | |
| 1 | The device has a built-in unique identifier. **[8259A]** | No | 0 | | | | ID.AM-1 |
| 2 | The device can interface with enterprise asset management systems. | Yes | 1 | | | | ID.AM-(1)(2), PR-DS-3 |
| 3 | The device can provide the organization sufficient visibility into its characteristics. | Yes | 1 | | | | ID.AM-(1)(2)(4) |
| 4 | The device or the device's manufacturer can inform the organization of all external software and services the device uses, such as software running on or dynamically downloaded from the cloud. | Yes | 1 | | | | DE.CM-8, PR.IP-1, PR.PT-3 |
| | VULNERABILITY MANAGEMENT | | | | | | |
| 5 | The manufacturer will provide patches or upgrades for all software and firmware throughout each device's lifespan. | Yes | 1 | | | | PR.IP-1 |
| 6 | The device either has its own secure built-in patch, upgrade, and configuration management capabilities, or can interface with enterprise vulnerability management systems with such capabilities. | Yes | 1 | | | | PR.IP(1)(3), PR.PT-3 |
| 7 | The device either supports the use of vulnerability scanners or provides built-in vulnerability identification and reporting capabilities. | Yes | 1 | | | | DE.CM-8 |
| | ACCESS MANAGEMENT | | | | | | |
| 8 | The device can uniquely identify each user, device, and process attempting to logically access it. | Yes | 1 | | | | PR.AC-(1)(7) |
| 9 | The device can conceal password characters from display when a person enters a password for a device, such as on a keyboard or touch | Yes | 1 | | | | PR.AC-7 |

Figure 2

The assessment process consists of determining compliance (with the *Expectation*), providing proof of compliance via validation point or tool, listing HIPAA Security Rule control type[21], and including additional comments (optional).

**COMPLIANCE**
Values are:

| Compliance | Value | Definition |
|---|---|---|
| **Yes** | 1 | The MIoT device complies with the Expectation |
| **No** | 0 | The MIoT device does not comply with the Expectation |
| **PL (Partial-LOW)** | 0.25 | Device compliance for Expectation is Limited (0%-25%) |
| **PM (Partial-MODERATE)** | 0.50 | Device compliance for Expectation Moderate (25%-50%) |
| **PH (Partial-HIGH)** | 0.75 | Device compliance for Expectation is High (50%-75%) |
| **Does Not Apply** | 0 | Expectation does not apply to the device (requires explanation) |
| **Alternate Approach** | 1 | An alternate approach is used to comply with the Expectation |
| **Unknown** | 0 | it is unknown if compliance with Expectation is possible |

**Note:** compliance which exceeds 75% of Expectation requirement(s) is deemed fully compliant.

Numerical values (0-1) are automatically added to the value column based on compliance statement. Values are summed and used to determine overall level of MIoT cybersecurity and privacy protection compliance (with NISTIR 8228 *Expectations*).

**PROOF-OF-COMPLIANCE (VALIDATION POINT/TOOL)**
A process, procedure or tool (manual or automated) which is used as auditable proof or evidence that the *Expectation* is being satisfied.

For example, for DEVICE SECURITY/Asset Management Expectation: *the device has a built-in unique identifier* the validation point may be the short statement[22] "device ID/SN can be read by IT Asset Management System" with a tool reference to *ABC Asset Management* program or application. Validation is to provide sufficient supplementary or complementary information proving that the *Expectation* is being met.

[21] HIPAA Security Rule control types are Administrative, Technical or Physical (Privacy protection)
[22] Short, concise statements are preferred for clarity and readability

**SECURITY CONTROL/TYPE**

This value directly references HIPAA's Security Rule which requires that security controls be categorized as *Administrative*, *Technical* or *Physical*. Most often, a security control has but a single categorization but there are instances where a control may apply to multiple categories. For example, establishing an operational incident-handling capability may be categorized as both an Administrative and Technical control.

## COMPLETION AND COMPLIANCE

Once the MiOT assessment questionnaire is complete its information is then used to compute individual requirement and overall compliance with NISTIR 8228 *Expectations* in both numerical and graphic (radar or spider chart) formats (Figure 3). This view allows evaluators to identify areas of weakness or vulnerability and to determine the level of cybersecurity and privacy protection when considering the use of a MiOT device in their network environment.

**IoT Device Expectations**

**NISTIR-8228 Compliance Summary**

(Internal Use Only)

Compliance Score (%)

**93.27%**

| KEY: #, Expectation Name, Goal, # Questions | | Questions whose answers are: Responses | | | | | | | | | Calc. | Pctg. |
|---|---|---|---|---|---|---|---|---|---|---|---|---|
| | | Y | N | PL | PM | PH | D | A | U | Total | Value | % |
| EXP1: Asset Management [4] | DeS | 3 | 1 | 0 | 0 | 0 | 0 | 0 | 0 | 4 | 3.00 | 75.0% |
| EXP2: Vulnerability Management [3] | DeS | 3 | 0 | 0 | 0 | 0 | 0 | 0 | 0 | 3 | 3.00 | 100.0% |
| EXP3: Access Management [7] | DeS | 7 | 0 | 0 | 0 | 0 | 0 | 0 | 0 | 7 | 7.00 | 100.0% |
| EXP4: Incident Detection [4] | DeS | 4 | 0 | 0 | 0 | 0 | 0 | 0 | 0 | 4 | 4.00 | 100.0% |
| EXP5: Data Protection [4] | DaS | 2 | 0 | 0 | 1 | 1 | 0 | 0 | 0 | 4 | 3.25 | 108.3% |
| EXP6: Dissasociated Data Management [1] | PRI | 1 | 0 | 0 | 0 | 0 | 0 | 0 | 0 | 1 | 1.00 | 100.0% |
| EXP7: Informed Decision Making [1] | PRI | 1 | 0 | 0 | 0 | 0 | 0 | 0 | 0 | 1 | 1.00 | 100.0% |
| EXP8: PII Processing Permissions Mgmt. [1] | PRI | 1 | 0 | 0 | 0 | 0 | 0 | 0 | 0 | 1 | 1.00 | 100.0% |
| EXP9: Information Flow Management [1] | PRI | 1 | 0 | 0 | 0 | 0 | 0 | 0 | 0 | 1 | 1.00 | 100.0% |
| | | **23** | **1** | **0** | **1** | **1** | **0** | **0** | **0** | **26** | **24.3** | **93.3%** |
| | | | | | | | | | | **26** | | |

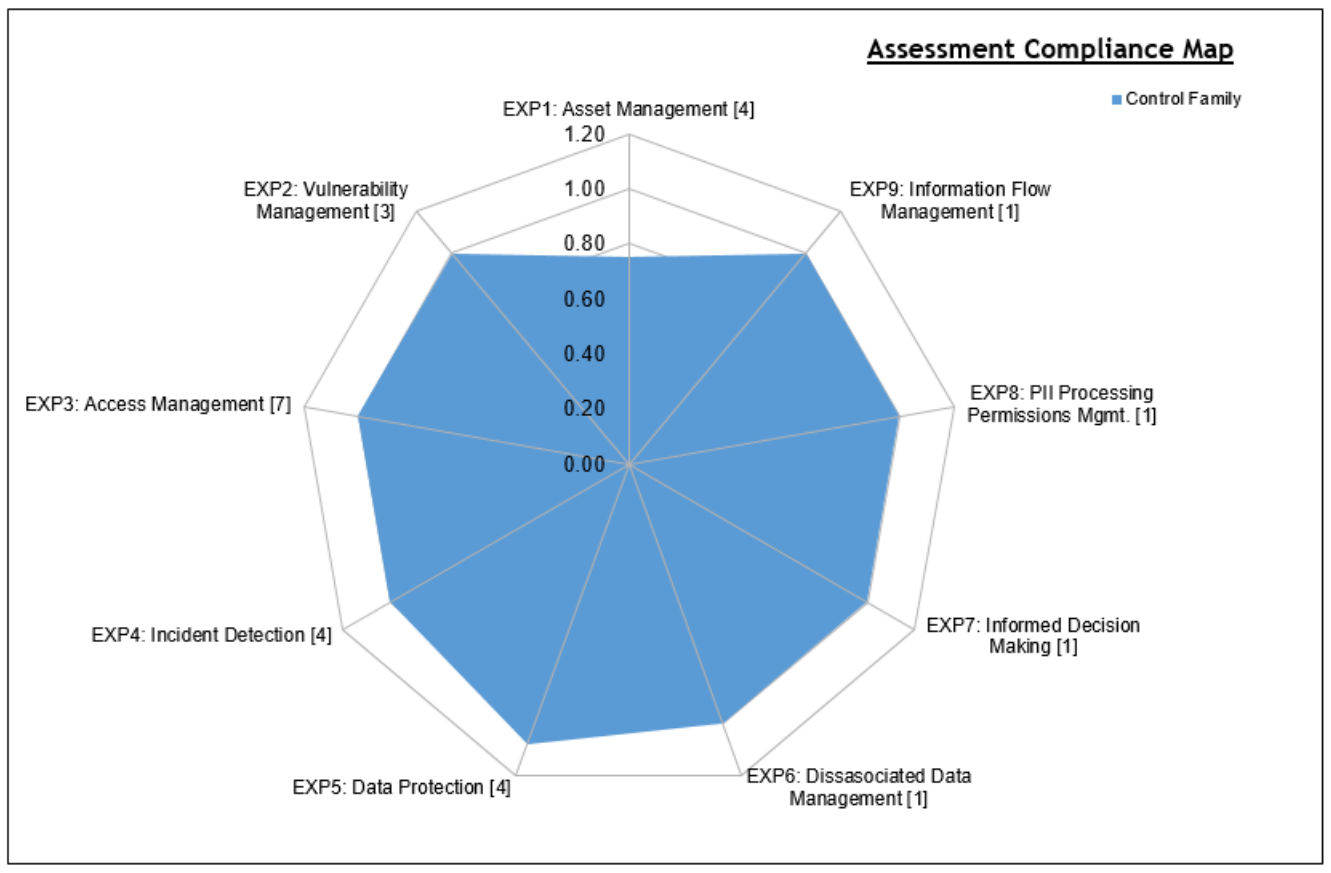

Figure 3

While only compliance results for NISTIR 8228 are shown in Figure 3 results for NISTIR 8228 & 8259A (*Cybersecurity recommendations for IoT manufacturers*) are also provided in the workbook. Distinction is made due to NISTIR 8259 series focus on device manufacturers instead of end-users of MiOT technology.

A data table and radar[23] (aka spider) chart provide tabular and graphical depiction of each *Expectation* value for aggregate compliance. The radar chart is especially useful for a birds-eye view (BEV) of deficiencies and areas which need to be addressed.

An acceptable compliance level is left to the discretion of the evaluator or organization as there is no published standard (although 80% or better is normally considered acceptable). Acceptable levels, however, can be designated for both individual *Expectation* and aggregate levels.

[23] A radar chart compares the values of three or more variables relative to a central point. It's useful when you cannot directly compare the variables and is especially great for visualizing performance analysis or survey data.

Regardless of threshold, compliance provides an organization with an idea of how well a MiOT device under consideration compares to established or recommended industry-standards. It is also used for determining the degree of risk (the MiOT represents) and ultimately whether or not its use is acceptable to a healthcare provider or organization.

**ADVANTAGES & BENEFITS OF USING NISTR 8228 FOR MiOT CYBERSECURITY ASSESSMENT**

1. NISTIR 8228 *Expectations* can be used to evaluate MiOT security in alignment with HIPAA\HITECH requirements. At present, there are few assessments designed specifically to evaluate MiOT cybersecurity risk and/or privacy protection.
2. Use of NISTIR 8228 *Expectations* take advantage of specific requirements for the protection of *Device*, *Data* and *Privacy* in order to evaluate risk.
3. Areas of weakness or vulnerability identified during the MiOT assessment can be used, in part, as the basis for risk determination and handling[24].
4. The security assessment approach offered through NISTIR 8228 is easy to use, flexible, repeatable, and employs current industry Best-Practices and guidance for both MiOT manufacturer and healthcare organization.
5. The use of NISTIR 8228 adheres to the intent of HITECH\'HIPAA Safe Harbor' provision which encourages and incentivizes healthcare providers and organizations to use NIST publications and Best-Practice guidance when considering Cybersecurity and Privacy Protection programs.
6. Employing NISTIR 8228 *Expectations* suggests an assessment process and methodology that is adaptable to non-federal sectors, and which can be used for regulatory and/or legal compliance.

**CONCLUSION**

Healthcare providers (i.e., Covered Entities) are mandated by HIPAA and HITECH to protect the Confidentiality, Integrity and Availability (CIA) of Electronic Protected Health Information (ePHI). This requirement extends to technology providers and Business Associates (BA) who directly support (healthcare) providers and use MiOT devices to satisfy this purpose.

NISTR 8228 can be used to evaluate MiOT device compliance with cybersecurity best practices relative to Device, Data and Privacy protection. Moreover, NISTIR 8228 associates its requirements with other NIST publications governing IoT and Cybersecurity including NIST Cybersecurity Framework, SP.800-53r5 (Security Control Catalog) and NISTIR 8259 series.

This paper has demonstrated how NIST guidance can be used to qualitatively evaluate the cybersecurity and privacy protection of MiOT devices based on level of compliance with specific NISTIR 8228 *Expectations*. Among other things, level-of-compliance aids in determining the cybersecurity and privacy protection risk an organization may face when considering the use of IoT-embedded medical devices. Such knowledge establishes "situational awareness" for how well a MiOT device aligns with legal & regulatory requirements, industry standards and best practices.

Finally, NISTIR-8228 can be used to create a documented history of compliance. Such history can serve as a guide for healthcare organizations when considering changes to its technology and/or network environment.

[24] Risk mitigation tiers are REDUCE, AVOID, ACCEPT and TRANSFER.